\documentclass[alpha-refs]{wiley-article}

\usepackage{siunitx}
\usepackage{amssymb,graphicx,placeins}
\usepackage{multirow}
\newcommand{\bY}{\mbox{\bf Y}}
\newcommand{\bX}{\mbox{\bf X}}

\newcommand{\bB}{\mbox{\bf B}}
\newcommand{\bA}{\mbox{\bf A}}
\newcommand{\bV}{\mbox{\bf V}}
\newcommand{\bT}{\mbox{\bf T}}
\newcommand{\bC}{\mbox{\bf C}}
\newcommand{\ba}{\mbox{\bf a}}
\newcommand{\bzero}{\mbox{\bf 0}}
\newcommand{\bSigma}{\mbox{\boldmath $\Sigma$}}
\newcommand{\bTheta}{\mbox{\boldmath $\Theta$}}
\newcommand{\bmu}{\mbox{\boldmath $\mu$}}

\papertype{Original Article}
\paperfield{Journal Section}

\title{A nonparametric approach to assess undergraduate performance}

\author[1]{Hildete P. Pinheiro PhD}
\author[2,3]{Pranab K. Sen PhD}
\author[1]{Alu\'{\i}sio Pinheiro PhD}
\author[1]{Samara F. Kiihl PhD}

\affil[1]{Department of Statistics, University of Campinas, Campinas, S\~ao Paulo, Brazil}
\affil[2]{Department of Biostatistics, University of North Carolina, Chapel Hill, NC, USA}
\affil[3]{Department of Statistics and Operations Research, University of North Carolina, Chapel Hill, NC, USA}

\corraddress{Hildete P. Pinheiro PhD, Department of Statistics, University of Campinas, Campinas, S\~ao Paulo, Brazil}
\corremail{hildete@g.unicamp.br}

\fundinginfo{CNPq-Brazil, Grant/Award Number: 308439/2014-7, 
308583/2015-9; 
Fapesp-Brazil, Grant/Award Number: 
2011/15047-7, 2013/00506-1, 2016/07226-2. 
}

\runningauthor{Pinheiro et al.}

\begin{document}

\maketitle

\begin{abstract}
Nonparametric methodologies are proposed to assess college students' performance. Emphasis is given
to gender and sector of High School. The application concerns the University of Campinas, a research university in Southeast Brazil. In Brazil college is based on a somewhat rigid set of subjects for each major. Thence a student's relative performance can not be accurately measured by the Grade Point Average or by any other single measure. We then define individual vectors of course grades. These vectors are used in pairwise comparisons of common subject grades for individuals that entered college in the same year. The relative college performances of any two students is compared to their relative performances on the Entrance Exam Score. A test based on generalized U-statistics is developed for homogeneity of some predefined groups. Asymptotic normality of the test statistic is true for both null and alternative hypotheses. Maximum power is attained by employing the union intersection principle. 

\end{abstract}

\section{Introduction}\label{intro}
The goal of this work is 
to evaluate the differences in students' performance from entrance to graduation in undergraduate courses. Some work has been done addressing this problem of performance assessment of undergraduate students.
 \cite{doi:10.1177/000494419103500105}, \cite{AEPA:AEPA274,AEPA:AEPA302} and \cite{AERE:AERE349} have studied the influence of students'  background and school factors on university performance, showing that students' success during their first year in the university is largely influenced by their EES (Entrance Exam Score) and type of high school (government or non-government). \cite{Murray-Harvey1993} used path analysis to identify characteristics of successful undergraduate students. \cite{SMITH2005549} and  \cite{e82d768146d940f299352c5a4d60cf37} have studied the schooling effects on university performance.
\cite{RePEc:oec:edukaa:5l4j85rq1s7b} used hierarchical regression models, with the {\it relative gain} as response variable, to investigate demographic and socio-economic factors that influenced 
university  performance. The {\it relative gain} is based on the relative rank of students' final (or last) recorded GPA (Grade Point Average) and students' total
EES rank. More recently, \cite{doi:10.1080/03610926.2013.804565,doi:10.1177/1471082X15596087} used binomial mixture models and quantile regression to model the number of credits gained by freshmen during the first year in college and \cite{doi:10.1080/02664763.2015.1077939} used nonparametric methods on
quasi U-statistics  \citep{PINHEIRO20091645,Pinheiro2011} to evaluate students' performance  in different groups. 

We present a real data set from the University of Campinas (Unicamp) in Section \ref{descriptive}. A descriptive analysis of the data set and the evaluation of students' performance based only on the  
overall average of the EES and the overall GPA is given. But the GPA might have some bias, since there are students from different areas with subjects/courses with different grading systems and different teachers. In view of this, we seek for more robust methods to compare students' performance.  
The proposed method looks at the EES rank and all the grades in all courses taken by each individual, performing all pairwise comparisons among individuals entering in the same year, in the same course/major, taking the same subject. 

 Average distance measures within and between groups for each year are defined and then the average over all years in the study is taken. The decomposability of quasi $U$-statistics \citep{PINHEIRO20091645,Pinheiro2011} is applied to
define average distance measures within and between groups. A test statistic for a
homogeneity test among groups is developed and its asymptotic
normality is achieved under the null hypothesis.  The alternative hypothesis is one-sided and, in order to maximize its power, we use the union intersection principal (UIP) discussed in \cite{SenBook} to test for contrasts of interest.   We study the performances of  Unicamp's students comparing them according to sex and type of High School - Public High Schools ({\it PuS}) and Private High Schools ({\it PrS}). The data set consists of all students who were admitted at Unicamp from 2000 to 2005.
Unicamp is a public institution,
located in the State of S\~ao Paulo and one of the top research universities in South America. Unicamp is highly selective, with an overall average  of over 15 candidates per
undergraduate position offered each year (www.comvest.unicamp.br). Public universities in Brazil are completely funded by the government with no tuitions or fees charged on the students. 

A detailed descriptive analysis of the data set is given in Section \ref{descriptive}. The notation and some basic results about U-statistics are presented in  Section 
\ref{notation}.  
The development of  a hypothesis test to evaluate  homogeneity between groups is given in Section \ref{hypothesis}.   Section \ref{application} illustrates an application of the test procedure with  the data set described in Section \ref{descriptive}. Finally, Section  \ref{discussion} presents a brief discussion of the results and methods.

\section{Undergraduate performance at the University of Campinas} \label{descriptive}

The dataset is composed by 12168  (57.3\% male and 42.7\% female) students which have enrolled at Unicamp at years 2000 to  2005 in Bachelor's degree courses/majors of the areas of Arts ({\it Ar}), Health Sciences ({\it HS}), Engineering and Exact Sciences ({\it EngES}) and Social Sciences ({\it SS}). Since in 2005 Unicamp implemented an affirmative action program giving extra bonus in the final EES for students who studied all High years in Public School, it was of great interest the study of performance of these students from entrance to graduation. In view of this, it would be interesting to have most of the students in the data set who already graduated from college. 
The academic situation of these students were classified as following: Graduates (students who have already graduated  - 77.1\%), Active (students who were still enrolled in the University at the time the data was provided and had not graduated yet - 0.9\%),  and Others (the ones who dropped out from the University - 22.0\%). 
The students were, in their majority, between 16 and 23 years old (94.3\%) from all Brazilian regions and enrolled in 45 different majors/courses from
the areas of {\it HS} (19.8\%), {\it EngES} (55.7\%), {\it SS} (18.5\%)
and {\it Ar} (6\%). About 70\%  of students who enrolled between 2000 and 2005 come
from Private High School ({\it PrS}). The groups of most interest in the analysis are sex and type of High School because of previous work \citep{RePEc:oec:edukaa:5l4j85rq1s7b, doi:10.1080/02664763.2015.1077939} with data set from Unicamp showing some differences in performance according to sex and type of High School. So, the distributions of sex and type of High School by year are shown in Tables \ref{sex} and \ref{Emed}, respectively.  Table \ref{EmedSex} shows the total number of students in each group of interest, i.e., according to type of High School and Sex.

\begin{table}[ht]
\caption{Gender distribution by year}\label{sex}
\begin{tabular}{c c c c c c c c c} \hline \hline
 \multirow{2}{*}{Sex} & \multicolumn{6}{c}{Entrance year} & \multicolumn{2}{c}{\multirow{2}{*}{Total}}  \\
             & \multicolumn{1}{c}{2000} & \multicolumn{1}{c}{2001} & \multicolumn{1}{c}{2002} & \multicolumn{1}{c}{2003} &  \multicolumn{1}{c}{2004} & \multicolumn{1}{c}{2005} & \\ \hline
             &  \% &  \% &  \% &  \% & \% & \% & n & \%\\ \hline \hline
            Male &    59.1    &     56.1    &    56.6    &    59. 4   & 55.9  &  56.6  & 6969    &   57.3 \\
            Female &    40.9    &    43.9    &   43.4    &     40.6    &   44.1 &  43.4  & 5199    &   42.7 \\ \hline
            Total   &   100.0   &   100.0   &   100.0   &    100.0   &  100.0 & 100.0    & 12168    &   100.0\\ \hline \hline
\end{tabular}
\end{table} 

\begin{table}[ht]
\caption{Distribution per year according to type of High School}\label{Emed}
\begin{tabular}{c c c c c c c c c} \hline \hline
 \multirow{2}{*}{High School} & \multicolumn{6}{c}{Entrance year} & \multicolumn{2}{c}{\multirow{2}{*}{Total}}  \\
             & \multicolumn{1}{c}{2000} & \multicolumn{1}{c}{2001} & \multicolumn{1}{c}{2002} & \multicolumn{1}{c}{2003} &  \multicolumn{1}{c}{2004} & \multicolumn{1}{c}{2005} & \\ \hline
             &  \% &  \% &  \% &  \% & \% & \% & n & \%\\ \hline \hline
            Private &    70.0    &     71.6    &    69.7    &    71. 0   & 73.0  &  67.1  & 8429    &   70.4 \\
            Public &    30.0    &      28.4    &   30.3    &     29.0    &   27.0 &  32.9  & 3543    &   29.6 \\ \hline
            Total   &   100.0   &   100.0   &   100.0   &    100.0   &  100.0 & 100.0    & 11972*    &   100.0\\ \hline \hline
            \multicolumn{9}{l}{ \footnotesize *There was no information about type of High School for 196 students.}
\end{tabular}
\end{table}

\begin{table}[ht]
\caption{Total number  of students by Type of High School and Sex}\label{EmedSex}
\begin{center}
\begin{tabular}{c c c} \hline \hline
Group & n & \% \\ \hline \hline
           Male - Private  & 4797 & 40\\ 
            Female - Private &    3632 &  30    \\
            Male - Public  &    2041    & 17 \\ 
            Female - Public    &   1502  &  13\\ \hline 
            Total & 11972* & 100 \\ \hline \hline
   \multicolumn{2}{l}{ \footnotesize *There was no information about type of High School for 196 students.}          
\end{tabular}
\end{center}
\end{table}



\begin{figure}[htb]
    \centering
        \includegraphics[scale= 0.7]{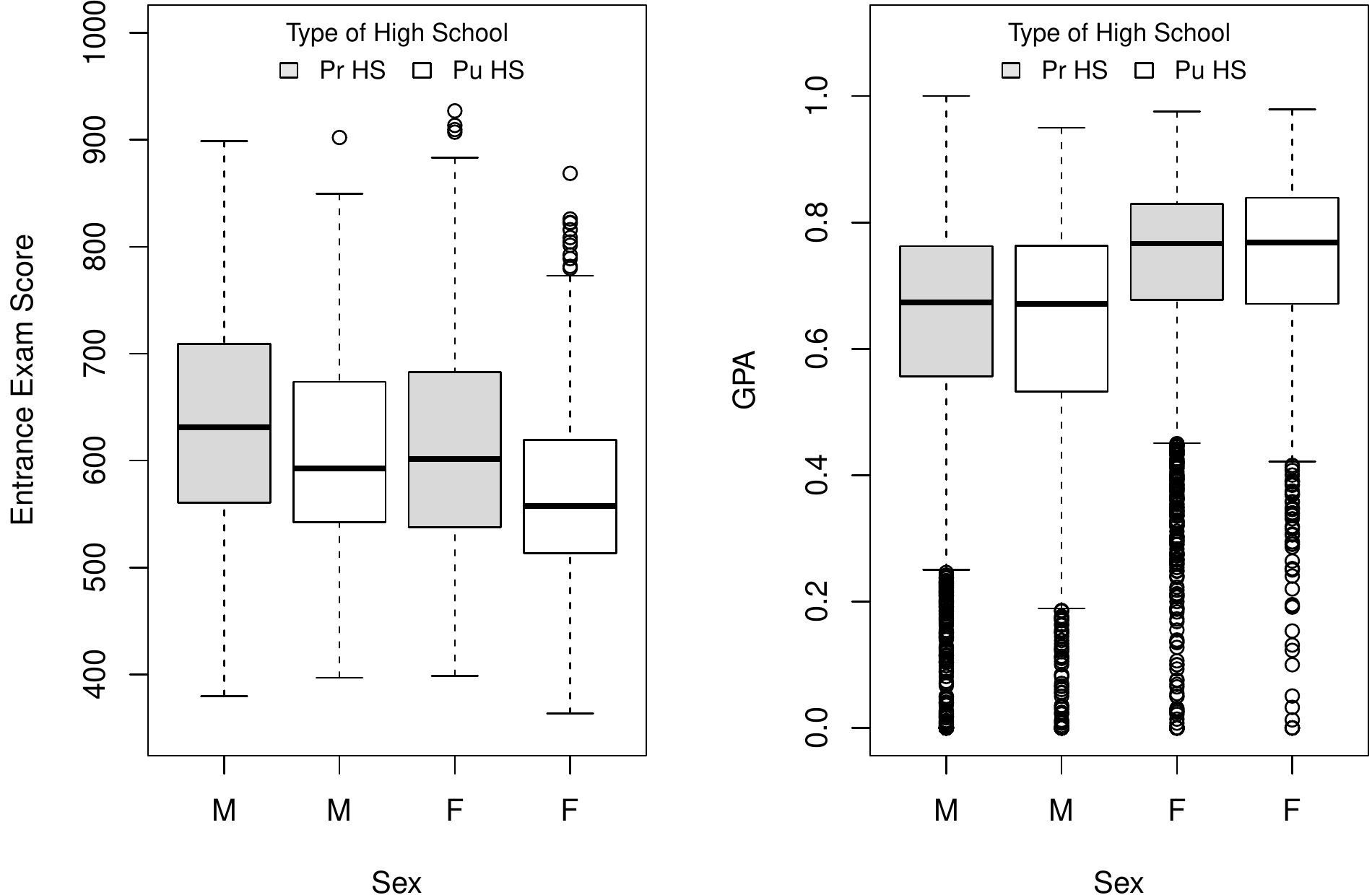}
        \caption{{\small Left:  Box plots of EES according to sex and type of High School. Right: Box plots of GPA  according to sex and type of High School }}
    \label{boxplotEES-GPA}
\end{figure}

\begin{figure}[htb]
    \centering
        \includegraphics[scale= 0.7]{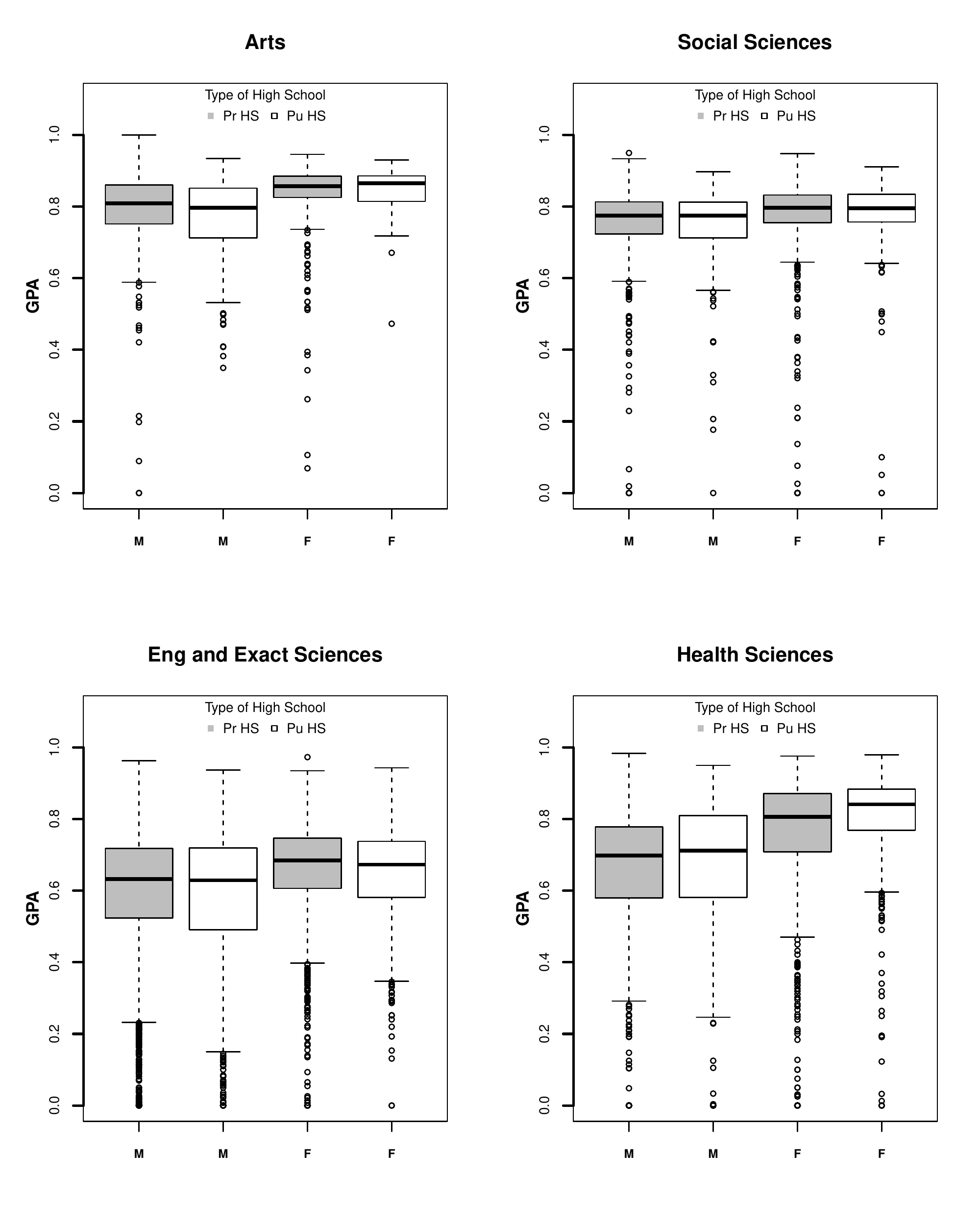}
        \caption{{\small Box plots of GPA for each area according to sex and type of High School. Arts 51\% males 49\% females, Social Sciences 37\% males and 63\% females, Engineering and Exact Sciences 70\% males and  30\% females, Health Sciences 41\% males and  59\% females.}}
    \label{boxplotEES-GPA-AREAS}
\end{figure}

Figure   \ref{boxplotEES-GPA} presents boxplots of sample distributions by gender and High School system of the EES (scores are standardized having mean 500 and standard deviation 100) and GPA (weighted average of grades from 0 to 10 according to the number of credits in each subject and it is between 0 and 1), respectively. One can see that  students who studied in {\it PrS} have a better EES performance than those coming from  {\it PuS} irrespective of gender. Once they get into College and we look at their GPA, the situation  seems to get reversed or at least, on average, they are tied. On the other hand, when one looks at the distribution of the GPA by sex, type of High School and Area displayed in Figure \ref{boxplotEES-GPA-AREAS}, the situation is not that clear and it is not the same in all areas.  The difference between sexes is not so big, especially in Exact Sciences and Engineering and in Social Sciences. In addition, there is a majority of Male students in Engineering and Exact Sciences (70\% males and  30\% females), while in Health Sciences (41\% males and  59\% females) and  Social Sciences (41\% males and  59\% females) the situation is reversed and in Arts (51\% males 49\% females) is quite even. The point is that students from different courses/majors take different subjects with different grading systems. Therefore, the GPA might not be a good measure of performance to compare students of different areas or majors.  
Furthermore, one can notice that there are many outliers at the lower tail of the distribution of the GPA scores. This is due to the 22\% of students who dropout from College. Some students who did not have a good performance in the first years of College (e.g., fail all the subjects in the first or second semester) have their enrollment canceled. Since Unicamp is a public university, there are some rigid dropout rules.  

 \FloatBarrier
 
 \FloatBarrier

  To investigate further the dropout rate at Unicamp, a logistic regression model was fitted having as response a dichotomous variable being Active/Graduated or dropout. The covariates in the model were Sex, Type of High School, Entrance year and Area. The significant main effects were Sex (p-value $<0.0001$), Area (p-value $<0.0001$) and Entrance Year (p-value $=0,0097$). The main effect for type of High School was not significant (p-value $=0.3468$), but there were significant interactions between Sex and type of High School (p-value $= 0.0041$) as well as Area and Entrance year (p-value$=0.0253$). According to the estimates of the parameters in the model, the highest rate of dropout are among Males, students from {\it EngES} and from the Entrance year of 2003. The interaction between Sex and type of High School  is explained as follows: within students coming from {\it PuS}, the estimated odds ratio is 2.22 (i.e, Male students have 2.2 times more chance of drop out than Female students), while within students coming from {\it PrS}, the estimated odds ratio is 1.59 (i.e., the dropout rate is 1.6 times higher for Male than Female students).  In general, the dropout rate is higher for Males than for Females, but the difference in the dropout rate is even higher for those coming from {\it PuS}. 

If the analysis were done just looking at the final EES and the GPA, probabilities of discordant and concordant pairs may be defined as the following. 
A probability of concordance of type 1 ($C_{1gg'}$) is when students of group $g$ are better than of those in $g'$ in the EES and  in the GPA,  discordance of type 1 ($D_{1gg'}$) is when students of group $g'$ are better than those in group $g$ in the EES, but worse than those of $g$ in the GPA,  concordance of type 2 ($C_{2gg'}$) is when students of group $g'$ are better than those of $g$ in the EES and in the GPA and discordance of type 2 ($D_{2gg'}$) is when students of group  $g$ are better than those of $g'$ in the EES, but worse than those of $g'$ in the GPA. So, if the probability of type I discordance is greater than the probability of type II, the student coming from group $g$ has a greater chance to  perform better in College than one coming from group $g'$. Table \ref{discpair} shows the proportion of concordance of type 1 ($C_{1gg'}$),   
discordance of type 1 ($D_{1gg'}$),   
concordance of type 2 ($C_{2gg'}$) 
and discordance of type 2 ($D_{2gg'}$).
The first two lines of Table \ref{discpair} show that women seems to be better in the EES and in the GPA than men (35\% and 36\% of the times they are better than men in {\it PuS} and {\it PrS}, respectively) against 28\% and 27\%, respectively, of the times that women are worse than men in the EES and better in in GPA. Line three  
 of Table \ref{discpair} shows that 31\% of Female  
 students from {\it PuS} were better in the EES and continue better in the GPA,  
 but there is not much difference from the case where Female students from {\it PrS} were better in the EES and continue better in the GPA (28\%). Looking at all lines of Table \ref{discpair}, there is not much difference among them, but the proportion of concordant pairs seems to be a little bigger than the discordant pairs.
 
\begin{table}[ht]
\caption{Proportion of discordant and concordant pairs by groups of interest according to final EES and GPA} \label{discpair}
\begin{center}
\begin{tabular}{c r r r r | r | r} \hline
Groups & $C_{1gg'}$ & $C_{2gg'}$ & $D_{1gg'}$ & $D_{2gg'}$ & $C$ & $D$ \\ \hline
$g=$F-PuS; $g'=$M-PuS & 0.35 & 0.22 & 0.28 & 0.15 & 0.57 & 0.43 \\
$g=$F-PrS; $g'=$M-PrS & 0.36 & 0.23 & 0.27 & 0.15 & 0.59 & 0.41\\
$g=$F-PuS; $g'=$F-PrS & 0.31 & 0.28 & 0.23 & 0.19 & 0.58 & 0.42\\
$g=$M-PuS; $g'=$M-PrS & 0.29 &0.29 & 0.22 & 0.19 & 0.59 & 0.41\\ \hline
\end{tabular}
\end{center}
\end{table}

Because of the potential bias of the GPA due to different areas and different types of grading system, we were motivated to seek for more robust methods to measure academic performance. Therefore, methods presented in sections \ref{notation} and \ref{hypothesis} perform all pairwise comparisons within students from the same course/major, same year of entrance and taking the same subject. 

\section{Notation and U-statistics} \label{notation}

Let ${\bf Z}_{iag} = (Z_{iag1}, \ldots , Z_{iagL_{i}})'$ be the vector of grades for student $i$, from group $g$, who entered at year $a$.   
Let $l=1,\ldots , L_i$  be the index indicating the subject taken by student $i$, with $L_i$ being all the subjects taken by student $i$. 
Also, let  $\bar{Z}_{0ig}$ and $\bar{Z}_{0jg}$ be the average of the EES for students $i$ and $j$, respectively.

Note that though $Z$'s are theoretically continuous r.v.'s, we observe discrete grades, since they are rounded by one decimal point. So, let $Y_{iagl}$ be the discrete grades for student $i$. For instance, $Y_{iagl}\in \{0.0,0.1,0.2, \ldots , 10.0\}$, i.e., 
$Y_{iagl}=0.0$, if $Z_{iagl}\in [0.0, 0.05)$,  $Y_{iagl}=0.1$, if $Z_{iagl}\in [0.05, 0.15)$, $\ldots$, 
$Y_{iagl}=9.9$, if $Z_{iagl}\in [9.85, 9.95)$, 
$Y_{iagl}=10.0$, if $Z_{iagl}\in [9.95, 10.0]$. 
Analogously, we can define $\bar{Y}_{0ig}$ and $\bar{Y}_{0jg}$ as the discrete versions of the EES  $\bar{Z}_{0ig}$ and $\bar{Z}_{0jg}$ for students $i$ and $j$, respectively.

Let $L_a=\sum_{i=1}^{n_a}L_i$ be the total number of subjects that students from year $a$ can take,  
 $L=\sum_{a=1}^A L_a$ be the total number of subjects that students from years $a=1, \ldots A$ can take, 
 $n_{agl}$ be the number of students of group $g$ from year $a$ who took subject $l$, $l=1, \ldots , L$,  
 $n_{ag}$ be the number of students of group $g$ from  year $a$ and $n_a (= \sum_{g=1}^G n_{ag})$ be the number of students in year $a$.  
 
Typically, $L$ is large, but each student takes a number of subjects (say 8 or 10) a year, so that 
$\sum_{g=1}^G \sum_{l=1}^{L} n_{agl} >> n_a$. 

 Also let 
$I\!\!\!I_{agg'l}(i,j) = I\!\!\!I_{agl}(i)\times I\!\!\!I_{ag'l}(j)$, where \\
$\left \{ \begin{array}{l} 
I\!\!\!I_{agl}(i) = I\!\!\!I(\mbox{student  $i$ of year  $a$ and group $g$  took subject  $l$}) 
\mbox{ and } \\
  I\!\!\!I_{ag'l}(j)= I\!\!\!I(\mbox{student  $j$ of year $a$ and group $g'$ took subject  $l$}),  \mbox{with $I(B)=1$, if $B$ is true and $0$ otherwise.}
  \end{array} \right .$ \\

Now define 
\begin{eqnarray}
 & &\!\!\!\!\!\!\phi(\bY_{iag}, \bY_{jag'}) \!=\! \phi(Y_{iagl}, Y_{jag'l}, \bar{Y}_{0ig}, \bar{Y}_{0jg'}) \!=\! \left \{I( Y_{iagl} > Y_{jag'l})I(\bar{Y}_{0ig} < \bar{Y}_{0jg'}) \right . \nonumber \\
 & & ~~ + I( Y_{iagl} < Y_{jag'l})I(\bar{Y}_{0ig} > \bar{Y}_{0jg'})  - I( Y_{iagl} > Y_{jag'l})I(\bar{Y}_{0ig} > \bar{Y}_{0jg'}) \nonumber \\ 
 & & ~~ -\left . I( Y_{iagl} < Y_{jag'l})I(\bar{Y}_{0ig} < \bar{Y}_{0jg'}) \right \} \left \{ I\!\!\!I_{agg'l}(i,j) \right \} \label{kernel}
\end{eqnarray}
as a kernel of a generalized $U$-statistics. 

Note that, according to (\ref{kernel}), there is no interest in tie cases, i.e., $Y_{ail}=Y_{ajl}$ or $\bar{Y}_{0i}=\bar{Y}_{0j}$.  Then,  $\phi(\bY_{iag}, \bY_{jag'})=0$ when there are ties. 

Now define a $U$-statistic \citep{10.2307/2235637, CIS-6492} of degree 2 as 
$U_{naggl} = \sum_{1\leq i<j\leq n_{agl}} 
\phi(\bY_{iag}, \bY_{jag})/
\binom{n_{agl}}{2}
$
and a generalized $U$-statistic of degree (1,1) as 
$U_{nagg'l} = \sum_{i=1}^{n_{agl}}\sum_{j=1}^{n_{ag'l}} 
\phi(\bY_{iag}, \bY_{jag'}) /
 n_{agl}n_{ag'l} $.

Now let the overall probability of concordance to be 
$P(C_{agg'l}) 
= P(C_{1agg'l}) + P(C_{2agg'l})$,  
where $P(C_{1agg'l}) =  P(Y_{iagl}>Y_{jag'l}, \bar{Y}_{0ig}>\bar{Y}_{0jg'})$ and  $P(C_{2agg'l}) = P(Y_{iagl}<Y_{jag'l}, \bar{Y}_{0ig}<\bar{Y}_{0jg'})$ are the probabilities 
of concordance of types 1 and 2, respectively.   Analogously, the overall probability of discordance is 
$P(D_{agg'l})  = P(D_{1agg'l}) + P(D_{2agg'l})$,
where $P(D_{1agg'l}) = P(Y_{iagl}>Y_{jag'l}, \bar{Y}_{0ig}<\bar{Y}_{0jg'})$ and 
$P(D_{2agg'l}) = P(Y_{iagl}<Y_{jag'l}, \bar{Y}_{0ig}>\bar{Y}_{0jg'})$ are 
the probabilities of discordance of types 1 and 2, respectively. 

Also, let
\begin{eqnarray}
& & \theta_{agg'l} = E\{\phi(Y_{iagl}, Y_{jag'l}, \bar{Y}_{0ig}, \bar{Y}_{0jg'} )\} =   P(D_{agg'l}) - P( C_{agg'l})   \label{theta-agg'l}
\end{eqnarray}
and
\begin{eqnarray}
& & \!\!\! \nu_{agg'l} = E\{[\phi(Y_{iagl}, Y_{jag'l}, \bar{Y}_{0ig}, \bar{Y}_{0jg'} )]^2 \} = P(D_{agg'l}) + P( C_{agg'l})  <1, \label{nu-agg'l}
\end{eqnarray}
since we are not interested in the ties. Analogously,
\begin{eqnarray}
& &  \theta_{aggl}  = E\{\phi(Y_{iagl}, Y_{jagl}, \bar{Y}_{0ig}, \bar{Y}_{0jg} )\} =   P(D_{aggl}) - P( C_{aggl}) \label{theta-aggl}
\end{eqnarray}
and
\begin{eqnarray}
& & \!\!\! \nu_{aggl} = E\{[\phi(Y_{iagl}, Y_{jagl}, \bar{Y}_{0ig}, \bar{Y}_{0jg} )]^2 \} = P(D_{aggl}) + P( C_{aggl})  <1 .\label{nu-aggl}
\end{eqnarray}

Using Hoeffding's decomposition \citep{10.2307/2235637}, we have 
\[ U_{naggl} =   \theta_{aggl} + \frac{2}{n_{agl}} \sum_{i=1}^{n_{agl}}[ \Psi_1(\bY_{iag}) -\theta_{aggl}]+ U_{naggl}^{*(2)} \]
and
\[ U_{nagg'l}  =   \theta_{agg'l} + \frac{1}{n_{agl}} \sum_{i=}^{n_{agl}} [\Psi_{10}(\bY_{iag}) - \theta_{aagg'l}] + 
\frac{1}{n_{ag'l}}\sum_{j=1}^{n_{ag'l}}[ \Psi_{01}(\bY_{jag'}) - \theta_{agg'l}] + U_{nagg'l}^{*(2)}, \]
where $\Psi_1(\bY_{iag}) = E[\phi(\bY_{iag}, \bY_{jag}) \mid \bY_{iag}]$,  
$U_{naggl}^{*(2)} = O_p(n_{agl}^{-1})$, \\
$\Psi_{10}(\bY_{iag}) = E[\phi(\bY_{iag},\bY_{jag'})\mid \bY_{iag}]$, 
$\Psi_{01}(\bY_{jag'}) = E[\phi(\bY_{iag}, \bY_{iag'})\mid \bY_{jag'}]$ and 
$U_{nagg'l}^{*(2)} = O_p( n_{agl}^{-1} + n_{ag'l}^{-1} + n_{agl}^{1/2}n_{ag'l}^{1/2})$. 

Since $U_{naggl}$ and $U_{nagg'l}$ are $U$-statistics, they are asymptotically normally distributed \citep{10.2307/2235637} as follows
\[ \sqrt{n_{agl}} (U_{naggl} - \theta_{aggl}) \stackrel{{\cal D}}{\longrightarrow} N(0, 4 \xi_1), \]
with $\xi_1 = E[\Psi_1^2(\bY_{iag})] - \theta_{aggl}^2$, and
\[ \gamma_n^{-1/2} (U_{nagg'l} - \theta_{agg'l})\stackrel{{\cal D}}{\longrightarrow} N(0, 1), \]
where $\gamma_n = \frac{\xi_{10}}{n_{agl}} + \frac{\xi_{01}}{n_{ag'l}}$, $\xi_{10} = E[\Psi_{10}^2(\bY_{iag})] - \theta_{agg'l}^2$ and $\xi_{01}= E[\Psi_1^2(\bY_{jag'})] - \theta_{agg'l}^2$. 

Now define a quasi $U$-statistics (Pinheiro et al., 2009; 2011) as 
\begin{equation}\label{Bagg'l}
 B_{nagg'l} = w_{nagg'l} \left \{ 2U_{nagg'l} - U_{naggl} - U_{nag'g'l} \right \},  
 \end{equation}
 with 
 \begin{equation}
 w_{nagg'l} = \frac{ \left [ \frac{ n_{agl}n_{ag'l} }{ n_{agl}+n_{ag'l} } \right ] }{
\left [  \sum_{g,g'}   \left (\frac{ n_{agl}n_{ag'l} }{ n_{agl}+n_{ag'l} } \right ) \right ]}.
\end{equation}
Note that 
\[ w_{nagg'l} = \left \{  \begin{array}{cl}
O(1), & \mbox{if both $n_{agl}$ and $n_{ag'l}$ are large} \\
O(n^{-1}), & \mbox{ if at least one of them ($n_{agl}$, $n_{ag'l}$) are small},  
\end{array} \right .
\]
where $n=\min\{n_{agl}, n_{ag'l}\}$. 

Now, let $B_{nagg'l} = B_{nagg'l}^{(1)} + B_{nagg'l}^{(2)}$, so that $B_{nagg'l}^{(1)}$ are those with $w_{nagg'l}= O(1)$ and $B_{nagg'l}^{(2)}$ are those with 
 $w_{nagg'l}=O(n^{-1})$.

Finally, let us define a test statistic as an overall sample measure of divergence between groups as   
\begin{eqnarray}
& & \!\!B_{ngg'} = \sum_{a=1}^A\sum_{l=1}^L  B_{nagg'l} =  \sum_{a=1}^{A}\left [ \sum_{l=1}^{L_a^{(1)}}B_{nagg'l}^{(1)} + \sum_{l=1}^{L_a^{(2)}}B_{nagg'l}^{(2)}\right ]  \nonumber \\
& & ~~~ = B_{ngg'}^{(1)} + B_{ngg'}^{(2)} \label{Bngg'}
\end{eqnarray}
 where $L_a^{(1)}$ is the total number of subjects with large $n_{agl}$'s and $n_{ag'l}$'s and $L_{a}^{(2)}$ is the total number of subjects with small $n_{agl}$'s and $n_{ag'l}$'s.  

Under the null hypothesis of homogeneity among groups, we would say that  there is no difference in performance between the entrance grade and the grades in the courses taken in the University neither within the groups nor between groups, i.e., 
$H_0: \theta_{aggl} = \theta_{ag'g'l} = \theta_{agg'l}$, $\forall \; 1\leq g< g'\leq G$, $\forall \; a=1, \ldots , A$ and $\forall \; l=1, \ldots , L$, or $H_0: 2\theta_{agg'l} - \theta_{aggl}-\theta_{ag'g'l} = 0$. 
But even though $-1\leq \theta_{agg'l} \leq 1$, there are several interesting situations under which $H_1: 2\theta_{agg'l} - \theta_{aggl}-\theta_{ag'g'l}> 0$ (see the Appendix for details of the one-sided alternative).

Note that for $G$ groups, we have $G(G-1)/2=G^{*}$ group comparisons. Therefore, we may define a $G^{*} \times 1$ vector $\bB_n = \bB_n^{(1)} + \bB_n^{(2)}$ with elements $B_{ngg'}$'s.

\section{Hypotheses and testing procedure} \label{hypothesis}

Concordant pairs are those where individual $i$ had a better/worse grade than $j$ in the entrance exam and continued to be better/worse than $j$ in the course grade in the University. 
Let's say that individual $i$ came from a Public High School and $j$ from a Private High School. A discordant pair of type I is when individual $i$ had a worse performance than $j$ in the EES, but a better performance than $j$ in his/her course grade. A discordant of type II is the opposite situation, i.e., $i$ had a better performance than $j$ in the EES, but a worse performance than $j$ in his/her course grade.  From this setup, we would say that if the probability of type I discordance is greater than the probability of type II, the student coming from a Public High School has a greater chance to perform better in the University than one coming from a Private High School. 

If there is no difference in performance between groups $g$ and $g'$, $H_0$ is true and 
$\theta_{aggl} = \theta_{ag'g'l} = \theta_{agg'l} 
= \theta_{al}$ and therefore 
$E\{\bar{B}_{nagg'l}\}=w_{nagg'l}[2\theta_{al}-\theta_{al} -\theta_{al}] =0$
and $H_1:  2\theta_{agg'l} - \theta_{aggl} - \theta_{ag'g'l} > 0$. Then, $E_{H_0}\{ B_{nagg'}   \}=0$ under $H_0$ and 
$E_{H_1}\{ B_{nagg'}   \}>0$. See the Appendix for justification of the one sided alternative. 


Note that the elements of $\bB_{n}$ are the ordered $B_{nagg'}$'s, say $B_{n1}, 
\ldots , B_{nG^*}$, where $G^{*} = \binom{G}{2}$ and they are not all independent. For instance, for $G=3$, $\bB_{n} = (B_{n12}, B_{n13}, B_{n23})$, with $G^{*} = \binom{3}{2}$.
Therefore, we need to make an adjustment like multiple comparisons test procedures. We may have different tests according to the interest. For instance, if we have two factors, say gender (F-Female and M-Male) and two types of High School (Pu-Public and Pr-Private), we may have statistics for the main effects and for the interaction. In this case, we have four groups (F-Pu, F-Pr, M-Pu, M-Pr) and for simplicity of notation $g=1, \ldots , 4$, with $1\rightarrow$ F-Pu, $2\rightarrow$ F-Pr, $3\rightarrow$ M-Pu, $4\rightarrow$ M-Pr. 

Then, we will have as  within and between graduate performance measures
\[ \left [\begin{array}{cccc}
\theta_{a11l} & \theta_{a12l} & \theta_{a13l} &\theta_{a14l} \\
                     & \theta_{a22l} & \theta_{a23l} & \theta_{a24l} \\
                      &                 & \theta_{a33l} & \theta_{a34l} \\
                      &                &                             & \theta_{a44l} 
                                                                       \end{array}  \right ] .
                                                                       \]
 
 In order to maximize the power of the tests, we may use the union intersection principle (UIP) discussed in \cite{SenBook}  to test for main effects of Sex and Type of High School as well as the interaction effect.

 For testing Female $\times$ Male: 
 \begin{eqnarray}
 H_{01}: 2 \theta_{a13l} - \theta_{a11l}-\theta_{a33l} + 2\theta_{a24l} - \theta_{a22l}-\theta_{a44l} =0 \nonumber \\
 H_{11}: 2 \theta_{a13l} - \theta_{a11l}-\theta_{a33l} + 2\theta_{a24l} -\theta_{a22l}-\theta_{a44l} >0 \label{uip1} 
 \end{eqnarray}
 and if we call $\Theta_{gg'} =2\theta_{agg'l} - \theta_{aggl} - \theta_{ag'g'l} $ the hypothesis can be written as 
 \begin{equation}
 H_{01}: \bC_1\bTheta=0 ~~~~~~~ \mbox {vs.} ~~~~~~
 H_{11}: \bC_1\bTheta>0,  \label{contrast1}
 \end{equation}
with $\bC_1 = (0, 1, 0, 0, 1, 0)$ and $\bTheta = (\Theta_{12}, \Theta_{13}, \Theta_{14}, \Theta_{23}, \Theta_{24}, \Theta_{34})'$. 

 For testing Public $\times$ Private:
 \begin{eqnarray}
 H_{02}: 2\theta_{a12l} + 2\theta_{a34l} -(\theta_{a11l}+\theta_{a22l}+\theta_{a33l}+\theta_{a44l}) =0 \nonumber \\
 H_{12}: 2\theta_{a12l} + 2\theta_{a34l} -(\theta_{a11l}+\theta_{a22l}+\theta_{a33l}+\theta_{a44l}) >0 \label{uip2}
 \end{eqnarray}
 or 
  \begin{equation}
 H_{02}: \bC_2\bTheta=0 ~~~~~~~ \mbox {vs.} ~~~~~~ 
 H_{12}: \bC_2\bTheta>0,  \label{contrast2}
 \end{equation}
with $\bC_2 = (1, 0, 0, 0, 0, 1)$ and $\bTheta = (\Theta_{12}, \Theta_{13}, \Theta_{14}, \Theta_{23}, \Theta_{24}, \Theta_{34})'$. 
  
  For testing the interaction Sex*Type of High School:
  \begin{eqnarray}
   H_{03}: 2\theta_{a12l} - \theta_{a11l}-\theta_{a22l} -2\theta_{a34l} +\theta_{a33l}+\theta_{a44l} =0 \nonumber \\
  H_{13}: 2\theta_{a12l} - \theta_{a11l}-\theta_{a22l} -2\theta_{a34l} +\theta_{a33l}+\theta_{a44l} \neq 0 \label{uip3}
   \end{eqnarray} 
   or 
    \begin{equation}
 H_{03}: \bC_3\bTheta=0 ~~~~~~~ \mbox {vs.} ~~~~~~
 H_{13}: \bC_3\bTheta \neq 0,  \label{contrast3}
 \end{equation}
with $\bC_3 = \left (  \begin{array}{cccccc}
1 & 0 & 0& 0 & 0 & -1 \\
0 & 1 & 0 & 0 &-1 & 0  \end{array} \right  )$ and $\bTheta = (\Theta_{12}, \Theta_{13}, \Theta_{14}, \Theta_{23}, \Theta_{24}, \Theta_{34})'$. 

Then we define a vector $\bB_n=(B_{n1}, \ldots , B_{nG^*})$ as the vector of the ordered $B_{ngg'}$'s. In this case, $\bB_n= (B_{n12}, B_{n13}, \ldots , B_{n34})$ is a $6\times 1$ vector. 
Also, $\bT_n = \bC\bB_n$ is the vector of linear combinations of the elements of $\bB_n= \bB_n^{(1)} + \bB_n^{(2)}$. Then, we may write $\bT_n= \bC\bB_n^{(1)} + \bC\bB_n^{(2)} = 
\bT_n^{(1)} + \bT_n^{(2)}$.

\begin{theorem} Let $\bC$ be a matrix of contrasts and $\bB_n^{(2)}$ be a vector with elements $B_{ngg'}^{(2)} =\sum_{a=1}^{A}\sum_{l=1}^{L_a^{(2)}} B_{nagg'l}^{(2)}$ with $w_{nagg'l} = O(n^{-1})$.
\[ \sqrt{n}\bC\bB_n^{(2)} \stackrel{{\cal L}_2}{\longrightarrow} \bzero . \]
 \end{theorem}

\noindent
Proof: \\
Note that the elements of $B_n^{(2)}$ are $B_{ngg'}^{(2)} =\sum_{a=1}^{A}\sum_{l=1}^{L_a^{(2)}} B_{nagg'l}^{(2)}$ with $w_{nagg'l} = O(n^{-1})$. \\
Then,  $B_{ngg'}^{(2)} = O_p(n^{-1})$ and $Var(\sqrt{n}B_{ngg'}^{(2)} ) = O(n^{-1}) \Rightarrow E( n || \bC\bB_n^{(2)} ||^2 ) \rightarrow 0$, i.e., $ \sqrt{n}\bC\bB_n^{(2)} \stackrel{{\cal L}_2}{\longrightarrow} \bzero$. 

\hfill{ } $\Box$

\begin{theorem} \label{chi2}
Let $\bC$ be a matrix of contrasts, $\bT_n= \bC\bB_n$ and $n\bB_n \stackrel{d}{\longrightarrow} N(\bzero, \bSigma)$. If $\bV_n \stackrel{p}{\longrightarrow} \bSigma_1$, with $\bSigma_1 = \bC\bSigma \bC'$, then $n^2\bT_n'\bV_n^{-}\bT_n \stackrel{d}{\longrightarrow} \chi^2_{[rank(\bV_n^{-})]}$. 
\end{theorem}

\noindent
Proof: \\
According to Pinheiro et al. (2009), the elements of $\bB_n$ are quasi U-statistics and it has an asymptotically multivariate normal distribution with covariance matrix $\bSigma$, i.e., 
$n\bB_n \stackrel{d}{\longrightarrow} N(\bTheta, \bSigma)$.  

As $\bT_n$ is a linear combination of a asymptotically multivariate normal vector, then  $n\bT_n \stackrel{d}{\longrightarrow} N(\bC\bTheta,\bSigma_1)$, where $\bSigma_1 = \bC\bSigma\bC'$, i.e., 
$n\bT_n \stackrel{d}{\longrightarrow} \bX$, where $\bX \sim N(\bmu, \bSigma_1)$, with $\bmu = \bC\bTheta$. 

By Searle (1971)(Theorem 2, page 57), if $\bX \sim N(\bmu, \bSigma_1)$, then $\bX'\bA\bX \sim \chi_{[rank(\bA), \bmu'\bA\bmu/2]}^2$ if and only if $\bA\bSigma_1$ is idempotent. 
Now, if $\bA=\bSigma_1^{-}$, then, $\bX'\bSigma_1^{-}\bX \sim \chi_{[rank(\bSigma_1^{-}), \bmu'\bSigma_1^{-}\bmu/2]}^2$. Therefore, under $H_0: \bmu=\bC\bTheta=\bzero$ and 
$\bX'\bSigma_1^{-}\bX \sim   \chi_{(rank(\bSigma_1^{-}))}^2$. 

If we now consider $\bV_n = \widehat{\bSigma_1}$ and if $\bV_n^{-}  \stackrel{p}{\longrightarrow} \bSigma_1^{-}$, we can say that 
$\bX'\bV_n^{-}\bX - \bX'\bSigma_1\bX = \bX'(\bV_n^{-} - \bSigma_1^{-})\bX = o_p(1)$. Also, $n^2\bT_n'\bV_n^{-}\bT_n - n^2\bT_n'\bSigma_1^{-}\bT_n = 
n^2\bT_n'(\bV_n^{-} - \bSigma_1^{-})\bT_n = o_p(1)$. 

As $n\bT_n  \stackrel{d}{\longrightarrow} \bX$ and $\bV_n^{-}  \stackrel{p}{\longrightarrow} \bSigma_1^{-}$, by Slutsky Theorem, $n^2\bT_n'\bV_n^{-}\bT_n - \bX'\bSigma_1^{-}\bX = o_p(1)$ and under $H_0$, $n^2\bT_n'\bV_n^{-}\bT_n  \stackrel{d}{\longrightarrow} \chi_{[rank(\bV_n^{-})]}^2$.

\hfill{ } $\Box$

Now, consider the set $\mathbb{P}$ of all $2^p$ vectors $\ba=(a_1, \ldots , a_p)'$, where $a_j$ can be either $0$ or $1$, and partition $\bT_n$ and 
 $\bV_n$ into $(\bT_{na}, \bT_{na'})$ and 
 \[
 \left (  \begin{array}{cc}
 \bV_{n a a} & \bV_{n a a'} \\
 \bV_{n a' a} & \bV_{n a' a'}  \end{array}\right )
 \]
 $\ba'$ being the complement of $\ba$, $\emptyset \subset \ba \subseteq \mathbb{P}$.  Further, define $\bT_{na:a'}$ and $\bV_{naa:a'}$ as 
 $\bT_{na:a'} = \bT_{na} - \bV_{naa'}\bV_{naa'}^{-} \bT_{na'}$ and $\bV_{naa:a'} = \bV_{naa} - \bV_{naa'}\bV_{na'a'}^{-}\bV_{na'a}, \forall \emptyset \subseteq \ba\subseteq \mathbb{P}$. Then,   
 for the hypotheses given in (\ref{contrast1}) and (\ref{contrast2}), we may use the union intersection principle \citep{SenBook} and  the test statistic is 
\begin{equation}
{\cal L}_n = \sum_{\emptyset \subseteq \ba \subseteq \mathbb{P}} 1\!\!1(\bT_{na:a'}>0, \bV_{na'a'}^{-}\bT_{na'}\leq \bzero)(n^{2}\bT_{na:a'}'\bV_{naa:a'}^{-}\bT_{na:a'}). 
\label{Ln}
\end{equation}
Under $H_0$, 
\begin{equation}
{\cal L}_n \stackrel{d}{\longrightarrow}\sum_{k=0}^p w_k \chi_k^2, \label{LnDist}
\end{equation}
where $\chi_k^2$ are independent chi-square random variables with $k(=0,1, \ldots, p)$ degrees of freedom  and the normal orthant probabilities with respect to $\bV_n$ lead to the approximation for the $w_k$ when sorted by the cardinality of the element $\ba: \emptyset \subseteq \ba \subseteq \mathbb{P}$.  

For the hypotheses given by (\ref{contrast3}) and Theorem \ref{chi2}, we have that, under $H_0$, 
\begin{equation}
{\cal L}_n = 
n^{2}\bT_{n}'\bV_{n}^{-}\bT_{n} \stackrel{d}{\longrightarrow} \chi_{\left [rank\left (\bV_n^{-}\right )\right ]}^2.
\label{Ln2}
\end{equation}

 If we are testing the difference in performance of students coming from Public or Private High School, with ${\cal L}_n$ one sided test, we will be able to detect in which direction is the difference, i.e., if students from $PuS$ have better performance in the University than students coming from $PrS$ or the other way around. 
 
  
We may use  $T_{n1}= B_{n13}+B_{n24}$, i.e, $\bC_1 = (0, 1, 0, 0, 1, 0)$ to test $H_{01}$, according to (\ref{contrast1}) with 
${\cal L}_{n1} = 1\!\!1(T_{n1}>0)n^2 T_{n1}^2/S_1^2$; 

 $T_{n2}= B_{n12}+B_{n34}$, i.e., $\bC_2=(1, 0, 0, 0, 0, 1)$ to test $H_{02}$, according to (\ref{contrast2}) with 
${\cal L}_{n2} = 1\!\!1(T_{n2}>0)n^2 T_{n2}^2/S_2^2$; and

In the case of a two-sided alternative,  \\
$\bT_{n3}=\left (\begin{array}{c} B_{n12} - B_{n34} \\
B_{n13}-B_{n24} \end{array} \right ) $, i.e., $\bC_3=\left (\begin{array}{cccccc} 1 & 0 & 0 & 0 & 0 & -1\\
                                                                                                                                        0 & 1 & 0 & 0 & -1 & 0 \end{array} \right )$ 
                                                                                                                                        can be used to test $H_{03}$, according to (\ref{contrast3}) with ${\cal L}_{n3} = n^2 \bT_{n3}'\bV_n^{-}\bT_{n3}$.

                                                                  

 \section{Application} \label{application}

We apply the proposed test procedures to the data from the University of Campinas described in Section \ref{descriptive}. 
The data set used to apply the methods shown in Sections \ref{notation} and \ref{hypothesis} is the same described in Section \ref{descriptive}. 
   
 The main interest is to test the following null hypotheses: \\
 $\bullet$ $H_{01}$: {\it There is no difference in performance between female and male}; \\
 $\bullet$  $H_{02}$: {\it There is no difference in performance between students coming from Public  and Private High  Schools}; \\
$\bullet$ $H_{03}$: {\it There is no interaction between sex and type of High School}; \\ addressed in Sections \ref{H1}, \ref{H2} and \ref{H3}, respectively.  

In order to apply the methods described in Sections \ref{notation} and  \ref{hypothesis}, we should separate the data into groups according to sex and type of High School. 
 
%

\subsection{Test of homogeneity between male and female students} \label{H1}

For testing Female $\times$ Male, the hypothesis test $H_{01}:\Theta_{13}+\Theta_{24}=0$ versus $H_{11}:\Theta_{13}+\Theta_{24}>0$ is given by  (\ref{contrast1}) with the test statistic given by 
$T_{n1} =B_{n13}+B_{n24}$, with ${\cal L}_{n1} = 1(T_{n1} > 0)n^2T_{n1}^2 /\sigma_1^2$. Since the p-value equals one, there is no evidence to reject the hypothesis of homogeneity between sexes. Note that the value of the observed test statistic ${\cal L}_{n1 obs}$ is 0, since we have a one-sided test and there is an indicator function in (\ref{Ln}).

\begin{figure}[htb]
  \centering
        \includegraphics[scale=0.8]{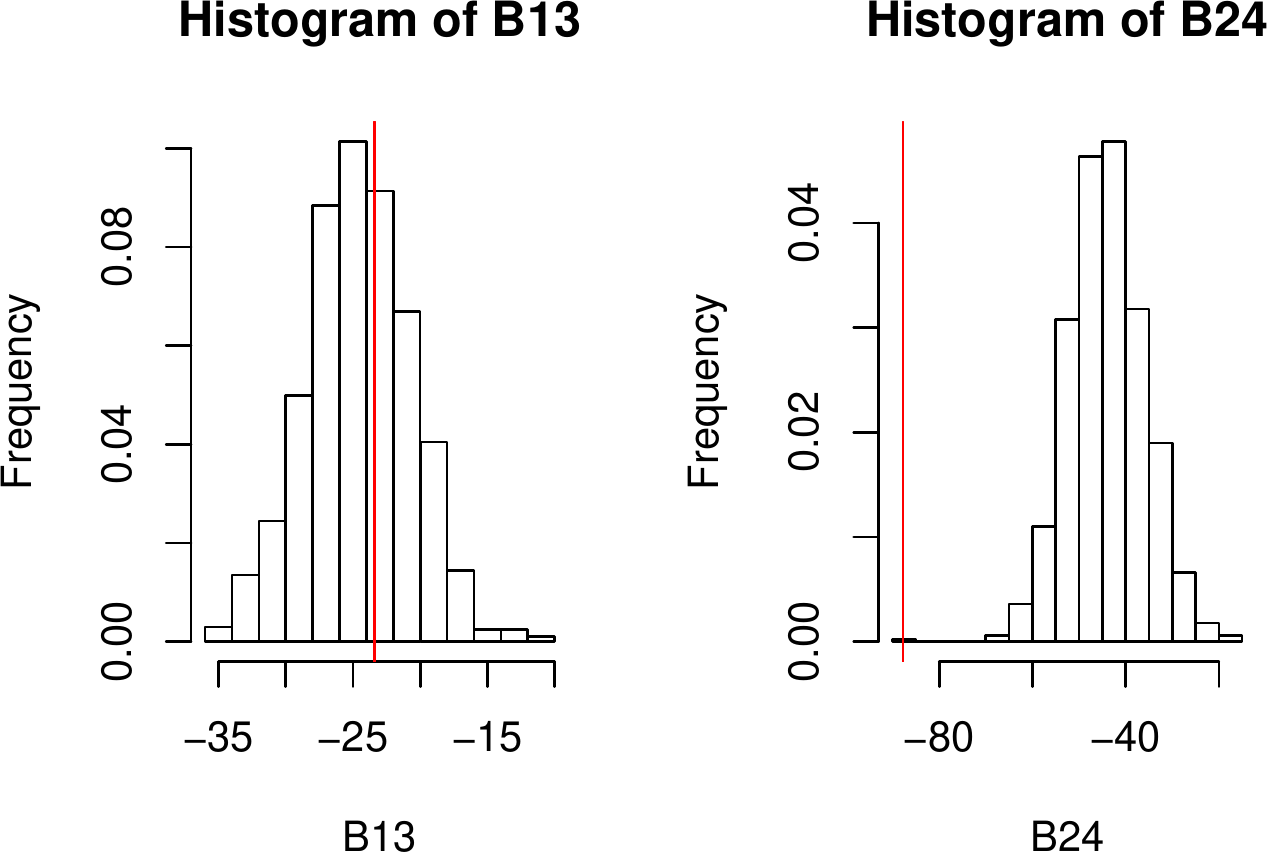}
        \caption{{\small Empirical distributions of $B_{n13}$ and $B_{n24}$ under the hypothesis of homogeneity between sexes.}}
    \label{HistSexos}
\end{figure}

Figure \ref{HistSexos} shows the empirical distribution of $B_{n13}$ (effect of sex in {\it PuS}, i.e., $H_0: \Theta_{13}=0$) and $B_{n24}$ (effect of sex in PrS, i.e, $H_0: \Theta_{24}=0$) under the null hypothesis. The value of the observed test statistics are $B_{n13obs}=-23.4$ and $B_{n24obs}=-87.75$ with p-values 0.387 and 1, respectively. Therefore, we can say that there is no evidence of changing of direction in performance for Female and Male students. 

 \FloatBarrier
 
\subsection{Test of homogeneity between students from public and private high schools} \label{H2}

\begin{figure}[htb]
  \centering
        \includegraphics[scale=0.8]{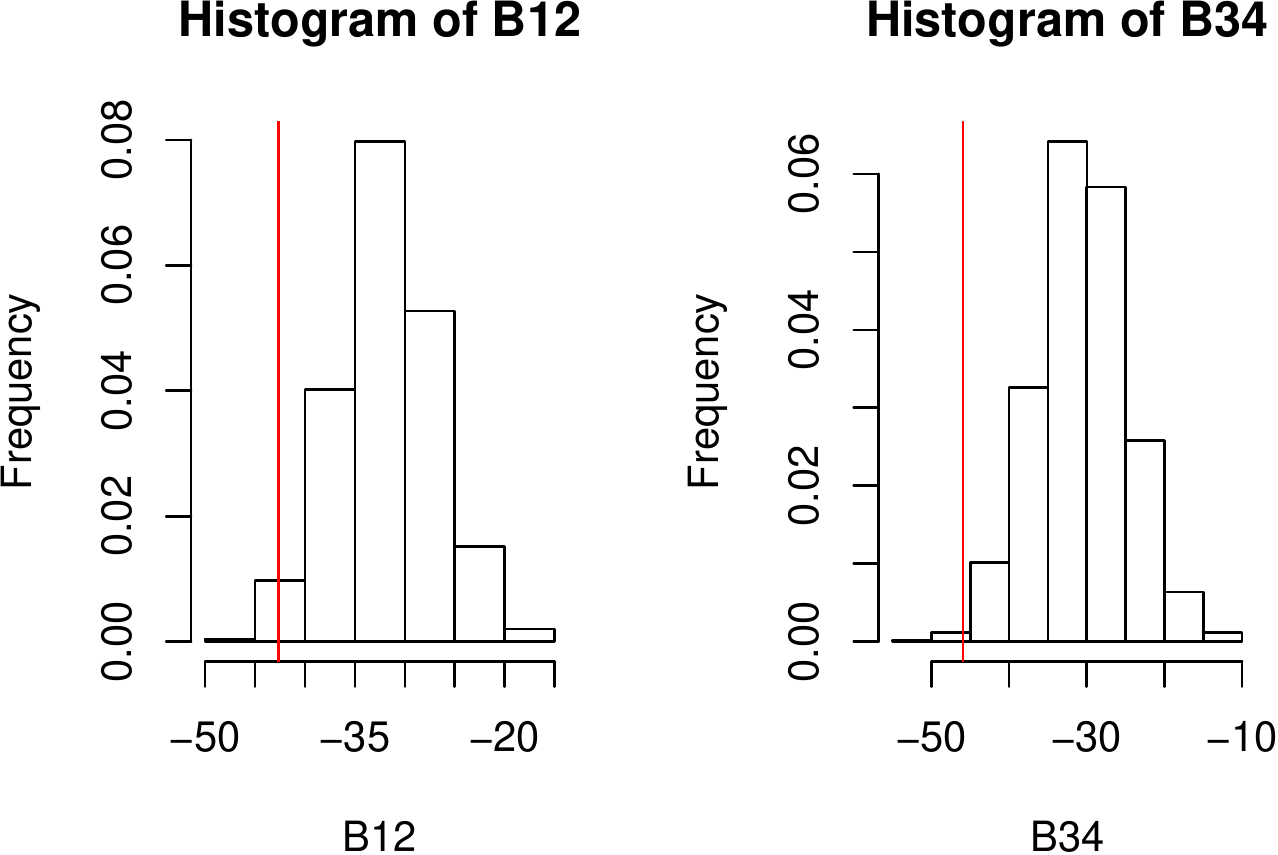}
        \caption{{\small Empirical distributions of $B_{n12}$ and $B_{n34}$ under the hypothesis of homogeneity between types of High School.}}
    \label{HistEscolas}
\end{figure}

For this test, we assume that students coming from {\it PrS} have better performance than those from {\it PuS}. So, we would like to test if the performance continues in the same direction during undergraduate school. The observed value of $B_{n12}$ and $B_{n34}$ are, respectively, -42.66 and -45.93, with p-values 0.990 and 0.995, respectively. Figure \ref{HistEscolas} presents the empirical distributions of $B_{n12}$ (effect  of High School among Males) and $B_{n34}$ (effect of High School among Females) under the null hypothesis of homogeneity among Schools. P-value for the test given by (\ref{contrast2}) and the test statistic (\ref{Ln2}) is 1, for the same reason given above for the test of homogeneity among sexes. From Figure \ref{HistEscolas} one can see that the effect of High School is similar in both sexes, with no evidence for rejecting the respective null hypotheses. So, we could say that the performance continues in the same direction, i.e., {\it PrS} students perform better than {\it PuS} students both in the EES and in College. 

 \FloatBarrier

\subsection{Test of interaction between sex and type of high school}\label{H3}

\begin{figure}[htb]
  \centering
        \includegraphics[scale=0.8]{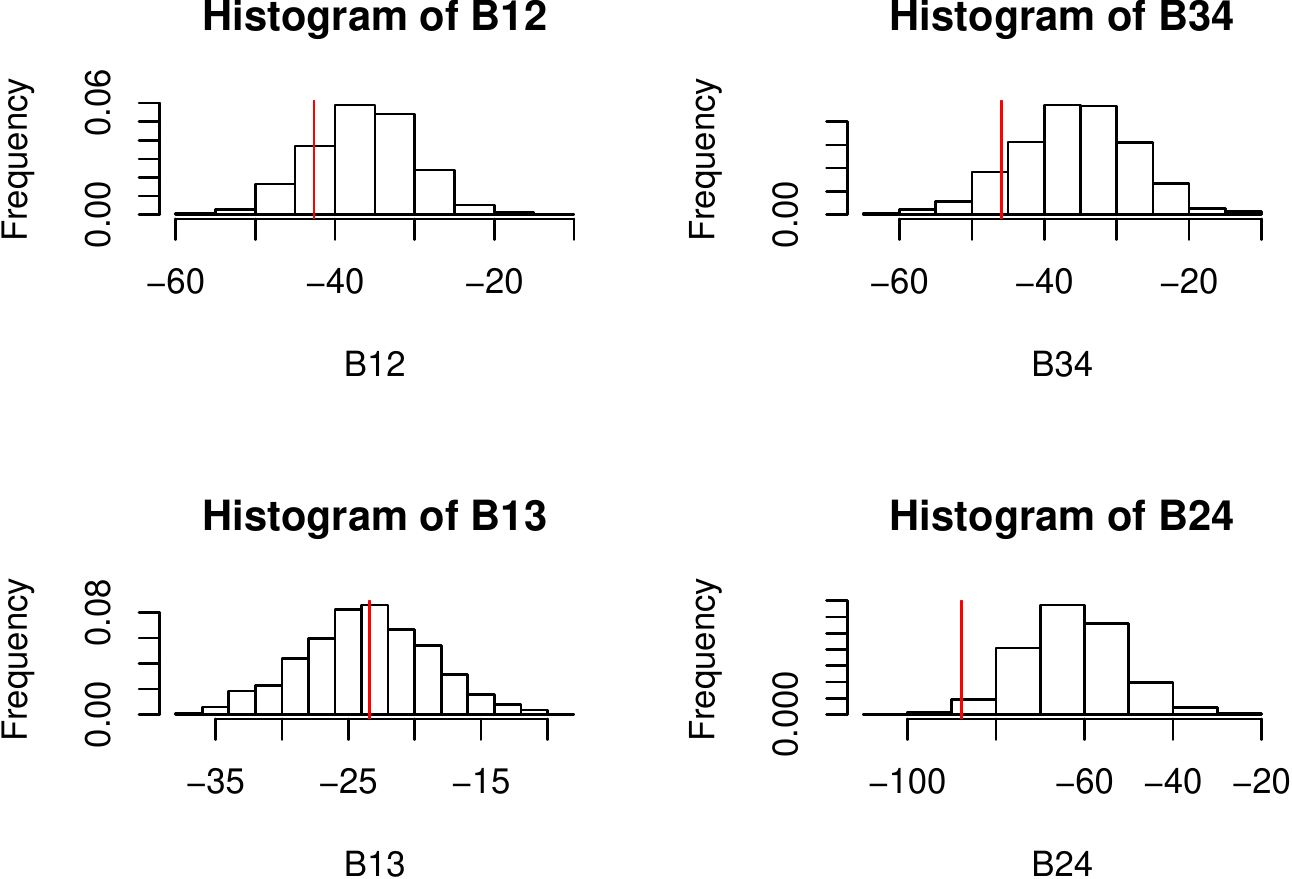}
        \caption{{\small Empirical distributions of $B_{n12}$ , $B_{n34}$ , $B_{n13}$ and $B_{n24}$  under the hypothesis of homogeneity between types of High School in both sexes.}}
    \label{HistInt}
\end{figure}

Figure \ref{HistInt} shows the empirical distribution of $B_{n12}$ (effect of High School system among Male students), $B_{n34}$ (effect of High School system among Female students), $B_{n13}$ (effect of sex among students from PuS)  and $B_{n24}$ (effect of sex among students from PrS) under the hypothesis of no interaction between type of High School and Sex. The values of the observed test statistics are $B_{n12obs} = -42.66$, $B_{n34obs} = -45.93$, $B_{n13obs} = -23.4$ and $B_{n24obs}=-87.75$. 

\begin{figure}[htb]
  \centering
        \includegraphics[scale=0.8]{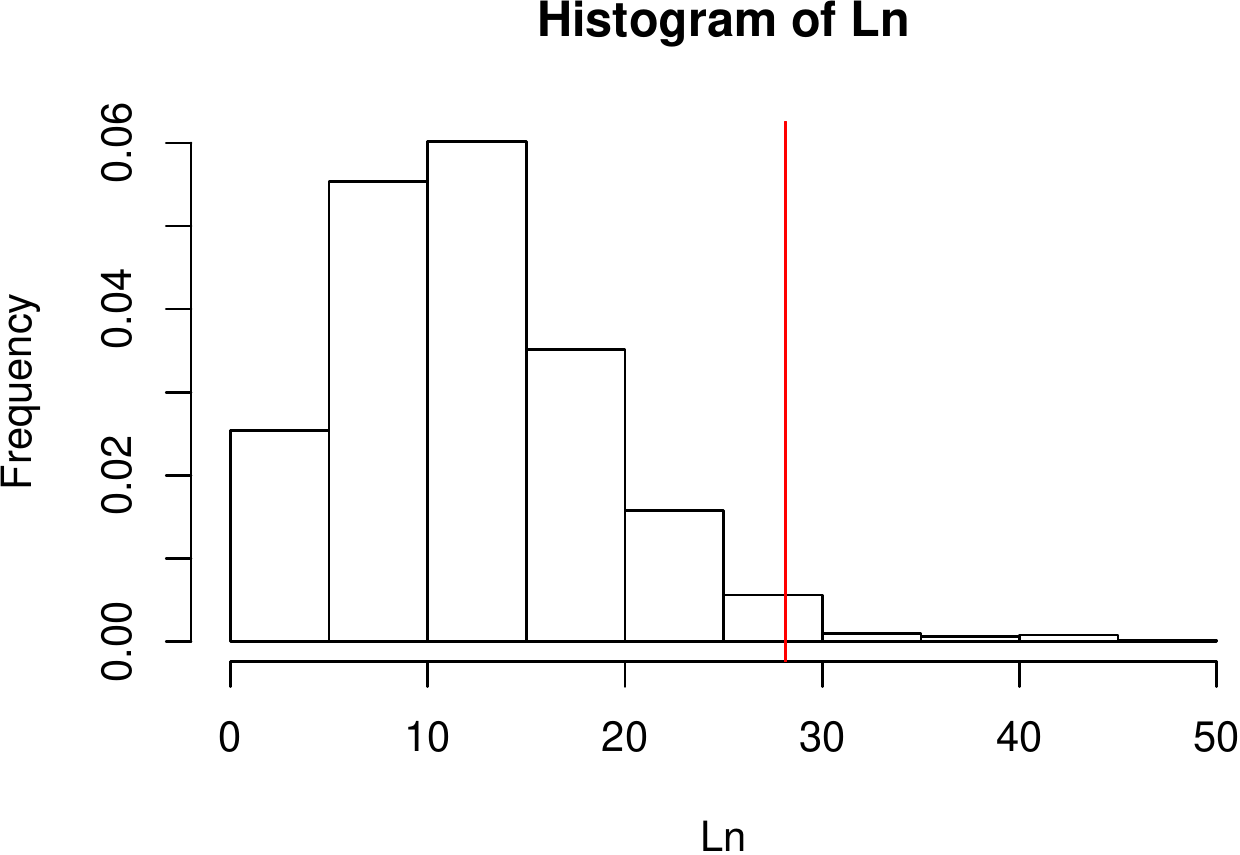}
        \caption{{\small Empirical distributions of $L_{n3}$  under the hypothesis of no interaction between types of High School and Sex.}}
    \label{HistLnH03}
\end{figure}

Figure \ref{HistLnH03} shows the empirical distribution of ${\cal L}_{n3}$ under the hypothesis of no interaction between sex and High School system. The observed value of the test statistic is ${\cal L}_{n3obs} = 28.135$, with p-value 0.02. Therefore, there is evidence of interaction between type of High School and Sex. Looking at Figure \ref{HistInt}, one can see a difference between the empirical distributions of $B_{n13}$ and $B_{n24}$, showing that the effects of sex in {\it PuS} ($B_{n13}$) and in {\it PrS} ($B_{n24}$) are different.  It seems that in {\it PrS} the difference between Female and Male students are greater than in {\it PuS}, which seems to agree with Table \ref{discpair}, where the number of concordant pairs in line two is slightly greater than in line one, indicating better performance of Female students in {\it PrS}. 

 \FloatBarrier

 \FloatBarrier


 
 \section{Discussion} \label{discussion}
 
We propose testing procedures for the comparison of student performance during college at the University of Campinas. These procedures are tailored for some  specificities common to Brazilian college course structure, but they are applicable to a wide range of problems in which shape restrictions and large random vectors play a role.   
The use of traditional measures like the GPA may mask the relative performance of a student. This is illustrated in Section \ref{descriptive}, and motivates the development of the proposed test statistic ${\cal L}_n$. 
The ${\cal L}_n$ statistic has other advantages, besides being designed to solve GPA shortcomings as an unbiased relative performance measure. 
The ${\cal L}_n$-based one sided test usually performs  better than the two-sided Hotteling's $\chi^2$-test. 
The  ${\cal L}_n$ test statistic is equivalent to the Hottelling's $T^2$ test when $\bT_n>\bzero$, but in situations where not all the components of $\bT_n$ are positive   ${\cal L}_n$ will be more powerful.
The problem of fairly assessing student performance is specially important when affirmative actions are employed. This is the case in several Brazilian universities, and these policies may greatly benefit from sound statistical analysis. We hope to be useful in this direction. 

  \section*{Appendix} \label{appendix}

  Let $\bar{Y}_{20}$ and $\bar{Y}_{10}$ be the EES from an individual of group 2 and an individual from group 1, respectively; 
  $Y_{2al}$ and $Y_{1al}$ are the grades in course $l$  of an individual from group 2 and an individual from group 1, respectively. 
  Let ${\rm Var}(\bar{Y}_{10}) = \sigma_1^2$, ${\rm Var}(\bar{Y}_{20l}) = \sigma_2^2$, ${\rm Var}(Y_{1al}) = \sigma_3^2$ and ${\rm Var}(Y_{2al}) = \sigma_4^2$. ${\rm Corr}(Y_{1al}, \bar{Y}_{10}) = \rho_1$ and ${\rm Corr}(Y_{2al}, \bar{Y}_{20}) = \rho_2$. Therefore, ${\rm Cov}(Y_{1al}, \bar{Y}_{10}) = \rho_1 \sigma_1 \sigma_3$ and 
  ${\rm Cov}(Y_{2al}, \bar{Y}_{20}) = \rho_2 \sigma_2 \sigma_4$. In general,
  \[
  \left (\begin{array}{c}   
   \bar{Y}_{20} - \bar{Y}_{10} \\
   Y_{2al} - Y_{1al}
   \end{array} \right ) 
   \sim N \left ( \left [ 
   \begin{array}{c} \mu_* \\
                              \mu_* 
      \end{array} \right ], 
     \left [ 
      \begin{array}{cc}
                      \sigma_1^2+\sigma_2^2 & \rho_1\sigma_1\sigma_3 +\rho_2\sigma_2\sigma_4 \\
                   \rho_1\sigma_1\sigma_3 +\rho_2\sigma_2\sigma_4   & \sigma_3^2 +\sigma_4^2 \end{array} \right ] \right  )                                                      
   \]

  We show some models for which the one-sided hypothesis is reasonable. 
  Suppose that, under $H_0$, students from group 2 do better than students from group 1 in the EES and the grades in undergraduate courses in group 2 will also be better than those from group 1, i.e, we expect more concordance than discordance.  Then, under $H_0$,
  $ \left \{ \begin{array}{l} 
  \bar{Y}_{20} \stackrel{\cal L}{=} \bar{Y}_{10} +\mu_*, \\
  Y_{2al} \stackrel{\cal L}{=} Y_{1al} + \mu_*  \end{array} \right .$,  with $\mu_*>0$. 
  
  Let $U= \bar{Y}_{20} -  \bar{Y}_{10}$ and $V= Y_{2al} - Y_{1al}$. 
  
  \noindent
  {\sf CASE A:} If  $\rho_1=\rho_2= \rho$  $H_0: \mu=0$ vs. $H_1: \mu>0$  \\
  Suppose $\sigma_1=\sigma_2$ and $\sigma_3=\sigma_4$, ${\rm Cov}(Y_{1al}, \bar{Y}_{10})= {\rm Cov}(Y_{2al}, \bar{Y}_{20}) = \rho\sigma_1\sigma_3$. 
  Then, ${\rm Cov}(U,V) = 2\rho\sigma_1\sigma_3$, ${\rm Corr}(U,V)=\rho$ and
   \[ \left ( \begin{array}{c} 
   U \\
   V
   \end{array} \right ) =
  \left (\begin{array}{c}   
   \bar{Y}_{20} - \bar{Y}_{10} \\
   Y_{2al} - Y_{1al}
   \end{array} \right ) 
   \sim N \left ( \left [ 
   \begin{array}{c} \mu_* \\
                              \mu_* 
      \end{array} \right ], 
     \left [ 
      \begin{array}{cc}
                      2\sigma_1^2 & 2\rho\sigma_1\sigma_3 \\
                    2\rho\sigma_1\sigma_3 & 2\sigma_3^2 \end{array} \right ] \right  ).                                                    
   \]
   Finally, let $Z_1 = (U-\mu_*)/\sqrt{2}\sigma_1$ and $Z_2=(V-\mu_*)/\sqrt{2}\sigma_3$. Then,  
    \begin{eqnarray}   
    P_{H_0}\{ {\rm Concordance} \} &\!\!=\!\!& P\left ( \frac{\bar{Y}_{20} - \bar{Y}_{10} }{\sqrt{2}\sigma_1}>0; \frac{ Y_{2al} - Y_{1al} }{\sqrt{2}\sigma_3} >0 \right )  
   \nonumber \\
& \!\!+\!\! &
 P\left (\frac{ \bar{Y}_{20} - \bar{Y}_{10}}{\sqrt{2}\sigma_1} <0; \frac{Y_{2al} - Y_{1al}}{\sqrt{2}\sigma_3} <0\right ) \nonumber \\
 &\!\!=\!\!& P ( Z_1 > -\mu_{1*}, Z_2 > -\mu_{2*}) + P (Z_1<-\mu_{1*}, Z_2<-\mu_{2*} ) \nonumber \\
  & \!\!=\!\!&  \int_{-\mu_{2*}}^{\infty} \int_{-\mu_{1*}}^{\infty} f_{Z_1,Z_2}(u,v) dudv +  \int_{-\infty}^{-\mu_{2*}} \int_{-\infty}^{-\mu_{1*}} f_{Z_1,Z_2}(u,v) dudv , \nonumber 
  \end{eqnarray}
  where $\mu_{1*} = \mu_*/(\sqrt{2}\sigma_1)$,  $\mu_{2*} = \mu_*/(\sqrt{2}\sigma_3)$ and 
  \begin{equation} \label{fz1z2}
 f_{Z_1,Z_2}(u,v)=  \frac{1}{2\pi \sqrt{1-\rho^2}} \exp \left \{ \frac{-1}{2(1-\rho^2)}\left [ u^2 - 2\rho u v + v^2\right ]\right \}.
    \end{equation}
  \begin{eqnarray}
  & & P_{H_0}\{ {\rm Discordance} \} = P( \bar{Y}_{20} - \bar{Y}_{10} >0; Y_{2al} - Y_{1al} <0)  \nonumber \\
  &\!\!+\!\! &P( \bar{Y}_{20} - \bar{Y}_{10} <0; Y_{2al} - Y_{1al} >0) \nonumber \\
  &\!\!=\!\!& P( Z_1 > -\mu_{1*}, Z_2 < -\mu_{2*} ) + P ( Z_1<-\mu_{1*}, Z_2>-\mu_{2*} ) \nonumber \\
  & \!\!=\!\!&  \int_{ -\infty}^{-\mu_{2*} }\int_{-\mu_{1*}}^{\infty} f_{Z_1,Z_2}(u,v) dudv +  \int_{-\mu_{2*}}^{\infty} \int_{-\infty}^{-\mu_{1*} } f_{Z_1,Z_2}(u,v) dudv. \nonumber 
  \end{eqnarray}
    Therefore,
    \begin{eqnarray}
   & & \theta_{12} = P_{H_0}\{ {\rm Discordance} \} - P_{H_0}\{ {\rm Concordance} \}  = \nonumber \\
  & \!\!=\!\!& \underbrace{  \int_{-\infty}^{-\mu_{2*}} \int_{-\mu_{1*}}^{\infty} f_{Z_1,Z_2}(u,v) dudv }_{(a)} + \underbrace{  \int_{-\mu_{2*}}^{\infty} \int_{-\infty}^{-\mu_{1*}} f_{Z_1,Z_2}(u,v) dudv}_{(b)} \nonumber \\
      & \!\!-\!\!&  \underbrace{ \int_{-\mu_{2*}}^{\infty} \int_{-\mu_{1*}}^{\infty} f_{Z_1,Z_2}(u,v) dudv}_{(c)} - \underbrace{  \int_{-\infty}^{-\mu_{2*}} \int_{-\infty}^{-\mu_{1*}} f_{Z_1,Z_2}(u,v) dudv }_{ (d) } \nonumber         
 \\    & \!\!=\!\!& (a) + (b) - (c) - (d) \mbox{   (under $H_0$)}  \label{theta12_H0}
      \end{eqnarray}

 Now, under $H_{1}:  \left \{ \begin{array}{l}
   \bar{Y}_{20} \stackrel{\cal L}{=} \bar{Y}_{10} + \mu_* ,\\
   Y_{2al} \stackrel{\cal L}{=} Y_{1al} + \mu_*-\mu$, ($\mu>0). \end{array} \right . $ 
 
    Then,
   \[
    \left ( \begin{array}{c} 
    \bar{Y}_{20} - \bar{Y}_{10} \\
    Y_{2al} - Y_{1al}  \end{array} \right )
   \sim N \left ( \left [ 
   \begin{array}{c} \mu_* \\
                              \mu_* - \mu 
      \end{array} \right ], 
     \left [ 
      \begin{array}{cc}
                      2\sigma_1^2 & 2\rho\sqrt{2}\sigma_1 \\
                   2\rho\sqrt{2}\sigma_3 & 2\sigma_3^2 \end{array} \right ] \right  )                                                      
   \]
 \begin{eqnarray}
  P_{H_1}\{ {\rm Concordance} \} &\!\!=\!\!&  
 P(Z_1 >-\mu_{1*}, Z_2 >  \mu_2 -\mu_{2*} ) + P(Z_1 < -\mu_{1*}, Z_2< \mu_2-\mu_{2*}  ) \nonumber \\
  & \!\!=\!\!&  \int_{\mu_2- \mu_{2*}}^{\infty} \int_{-\mu_{1*}}^{\infty} f_{Z_1,Z_2}(u,v) dudv +  \int_{-\infty}^{\mu_2-\mu_{2*}} \int_{-\infty}^{-\mu_{1*}} f_{Z_1,Z_2}(u,v) dudv \nonumber 
   \\
   & \!\!=\!\!&  \int_{-\mu_{2*}}^{\infty} \int_{-\mu_{1*}}^{\infty} f_{Z_1,Z_2}(u,v) dudv -  \int_{-\mu_{2*}}^{\mu_2-\mu_{2*}} \int_{-\mu_{1*}}^{\infty} f_{Z_1,Z_2}(u,v) dudv \nonumber \\
   & \!\!+\!\!&  \int_{-\infty}^{-\mu_{2*}} \int_{-\infty}^{-\mu_{1*}} f_{Z_1,Z_2}(u,v) dudv +  \int_{-\mu_{2*}}^{\mu_2-\mu_{2*}} \int_{-\infty}^{\mu_{1*}} f_{Z_1,Z_2}(u,v) dudv,   \nonumber 
  \end{eqnarray}
  where $\mu_{1*} = \mu_*/(\sqrt{2}\sigma_1)$, $\mu_{2*} = \mu_*/(\sqrt{2}\sigma_3)$ and $\mu_2 = \mu/(\sqrt{2}\sigma_3)$. 
   \begin{eqnarray}
  & & P_{H_1}\{ {\rm Discordance} \} =  P(Z_1 >-\mu_{1*}, Z_2 < \mu_2-\mu_{2*}  ) + P(Z_1 < -\mu_{1*}, Z_2> \mu_2-\mu_{2*}  ) \nonumber \\
  & \!\!=\!\!&  \int_{-\infty}^{\mu_2-\mu_{2*}} \int_{-\mu_{1*}}^{\infty} f_{Z_1,Z_2}(u,v) dudv +  \int_{\mu_2-\mu_{2*}}^{\infty} \int_{-\infty}^{-\mu_{1*}} f_{Z_1,Z_2}(u,v) dudv  \nonumber 
  \\
  & \!\!=\!\!&  \int_{-\infty}^{-\mu_{2*}} \int_{-\mu_{1*}}^{\infty} f_{Z_1,Z_2}(u,v) dudv +  \int_{-\mu_{2*}}^{\mu_2-\mu_{2*}} \int_{-\mu_{1*}}^{\infty} f_{Z_1,Z_2}(u,v) dudv \nonumber \\
   & \!\!+\!\!&  \int_{-\mu_{2*}}^{\infty} \int_{-\infty}^{-\mu_{1*}} f_{Z_1,Z_2}(u,v) dudv -  \int_{-\mu_{2*}}^{\mu_2-\mu_{2*}} \int_{-\infty}^{-\mu_{1*}} f_{Z_1,Z_2}(u,v) dudv \nonumber 
  \end{eqnarray}

   \begin{eqnarray}
   & & \theta_{12} = P_{H_1}\{ {\rm Discordance} \} - P_{H_1}\{ {\rm Concordance} \}  = \nonumber \\
  & \!\!=\!\!& \underbrace{  \int_{-\infty}^{-\mu_{2*}} \int_{-\mu_{1*}}^{\infty} f_{Z_1,Z_2}(u,v) dudv  }_{(a)}+ \underbrace{ \int_{-\mu_{2*}}^{\mu_2-\mu_{2*}} \int_{-\mu_{1*}}^{\infty} f_{Z_1,Z_2}(u,v) dudv}_{(e)} \nonumber \\
   & \!\!+\!\!&  \underbrace{ \int_{-\mu_{2*}}^{\infty} \int_{-\infty}^{-\mu_{1*}} f_{Z_1,Z_2}(u,v) dudv }_{(b)} -  \underbrace{ \int_{-\mu_{2*}}^{\mu_2-\mu_{2*}} \int_{-\infty}^{-\mu_{1*}} f_{Z_1,Z_2}(u,v) dudv}_{(f)} \nonumber \\
   & \!\!-\!\!& \underbrace{ \int_{-\mu_{2*}}^{\infty} \int_{-\mu_{1*}}^{\infty} f_{Z_1,Z_2}(u,v) dudv}_{(c)} + \underbrace{ \int_{-\mu_{2*}}^{\mu_2-\mu_{2*}} \int_{-\mu_{1*}}^{\infty} f_{Z_1,Z_2}(u,v) dudv}_{(e)} \nonumber \\
   & \!\!-\!\!& \underbrace{  \int_{-\infty}^{-\mu_{2*}} \int_{-\infty}^{-\mu_{1*}} f_{Z_1,Z_2}(u,v) dudv }_{(d)}- \underbrace{ \int_{-\mu_{2*}}^{\mu_2-\mu_{2*}} \int_{-\infty}^{-\mu_{1*}} f_{Z_1,Z_2}(u,v) dudv }_{(f)} \nonumber \\
    & \!\!=\!\!& (a) + (b)  - (c) - (d) +2(e) -2(f) \mbox{  (under $H_1$) } \label{theta12_H1}
      \end{eqnarray}
  Note that if $\rho>0$ and $f_{Z_1,Z_2}(u,v)$ is given by (\ref{fz1z2}), the expression given in (\ref{theta12_H1})$ > $ (\ref{theta12_H0}). Therefore, under $H_1: 2\theta_{12} >\theta_{11} +\theta_{22}$.  
   $\Box$   
  \\   
  \\     
%
\\
\\
{\sf CASE B:} $H_0: \rho_1 = \rho_2$ vs. $H_1: \rho_1<\rho_2$

 Now,   let $U_g =  \bar{Y}_{g0j} - \bar{Y}_{g0i}$,  $V_g = Y_{gajl} - Y_{gail}$, $U_{gg'} =  \bar{Y}_{g'0j} - \bar{Y}_{g0i}$,  $V_{gg'} = Y_{g'ajl} - Y_{gail}$, for $1\leq g< g'\leq G$. 
  
 Note that in general, ${\rm Var}(\bar{Y}_{g0i}-\bar{Y}_{g'0j}) = \sigma_1^2 +\sigma_2^2 = \sigma_{1*}^2$, 
  ${\rm Var}(Y_{gail}-Y_{g'ajl}) = \sigma_3^2 +\sigma_4^2 = \sigma_{2*}^2$, 
  ${\rm Cov}(\bar{Y}_{g0i}, Y_{gail})=\rho_1\sigma_1\sigma_3$ and  ${\rm Cov}(\bar{Y}_{g'0j}, Y_{g'jl})=\rho_2\sigma_2\sigma_4$. When comparing individuals from the same group, we will assume that $\sigma_1=\sigma_2$ and $\sigma_3=\sigma_4$.Then,
  
  \[
  \left (\begin{array}{c}   
  U_{1}\\ V_{1} \end{array}\right ) 
   \sim N \left ( \left [ 
   \begin{array}{c} \mu_* \\
                              \mu_* 
      \end{array} \right ], 
     \left [ 
      \begin{array}{cc}
                     2 \sigma_1^2 & \sigma_1\sigma_3(\rho_1+ \rho_2)\\
                     \sigma_1\sigma_3(\rho_1+ \rho_2) & 2\sigma_3^2 \end{array} \right ] \right  )                                                      
   \]
and
   \[
  \left (\begin{array}{c}   
  U_{12}\\ V_{12} \end{array}\right ) 
   \sim N \left ( \left [ 
   \begin{array}{c} \mu_* \\
                              \mu_* 
      \end{array} \right ], 
     \left [ 
      \begin{array}{cc}
                      \sigma_1^2+\sigma_2^2 & \rho_1\sigma_1\sigma_3 + \rho_2\sigma_2\sigma_4 \\
                    \rho_1\sigma_1\sigma_3 + \rho_2\sigma_2\sigma_4 & \sigma_3^2+\sigma_4^2 \end{array} \right ] \right  )                                                      
   \]
Definig $U_{1*} = U_1/(\sqrt{2}\sigma_1)$, 
$V_{1*}=V_1/(\sqrt{2}\sigma_3)$, 
$U_{12*} = U_{12}/\sqrt{ \sigma_1^2+\sigma_2^2}$, 
$V_{12*} = V_{12}/\sqrt{ \sigma_3^2+\sigma_4^2}$, $\mu_{1*}=\mu_*/(\sqrt{2}\sigma_1)$, $\mu_{2*}=\mu_*/(\sqrt{2}\sigma_3)$ we get  
  \[
  \left (\begin{array}{c}   
  U_{1*}\\ V_{1*} \end{array}\right ) 
   \sim N \left ( \left [ 
   \begin{array}{c} \mu_{1*} \\
                              \mu_{2*} 
      \end{array} \right ], 
     \left [ 
      \begin{array}{cc}
                     1 &(\rho_1+ \rho_2)/2\\
                     (\rho_1+ \rho_2)/2 & 1 \end{array} \right ] \right  )                                                      
   \]
and
   \[
  \left (\begin{array}{c}   
  U_{12*}\\ V_{12*} \end{array}\right ) 
   \sim N \left ( \left [ 
   \begin{array}{c} \mu_* \\
                              \mu_* 
      \end{array} \right ], 
     \left [ 
      \begin{array}{cc}
                     1& \frac{\rho_1\sigma_1\sigma_3 + \rho_2\sigma_2\sigma_4}{\sqrt{(\sigma_1^2+\sigma_2^2)(\sigma_3^2+\sigma_4^2)}} \\
                     \frac{\rho_1\sigma_1\sigma_3 + \rho_2\sigma_2\sigma_4}{\sqrt{(\sigma_1^2+\sigma_2^2)(\sigma_3^2+\sigma_4^2)}} &1 \end{array} \right ] \right  )                                                      
   \] 
  For $W_1=V_{1*}-\rho_{1*} U_{1*}$, ${\rm Cov}(W_1,U_{1*})=0$, 
  $W_1\sim N(\mu_{2*}-\rho_{1*}\mu_{1*}, 1-\rho_{1*}^2)$, with $\rho_{1*} = (\rho_1+\rho_2)/2$.  \\
  For $W_{12} = V_{12*}-\rho_{12*} U_{12*}$, ${\rm Cov}(W_{12}, U_{12*})=0$, with $\rho_{12*} = \frac{\rho_1\sigma_1\sigma_3 + \rho_2\sigma_2\sigma_4}{\sqrt{(\sigma_1^2+\sigma_2^2)(\sigma_3^2+\sigma_4^2)}}$. Then,

  \begin{eqnarray*}
   P(U_1>0, V_1>0) & = & P(U_{1*}>0, V_{1*}>0 )
    =  P(U_{1*}>0, W_1+\rho_{1*} U_{1*} > 0) \\
  &=& P( U_{1*}> 0, Z_{3*}> -\mu_{2*} +\rho_{1*}\mu_{1*}-\rho_{1*} u ) \\
  & = & \int_{0}^{\infty} [1 - \Phi( -\mu_{2*} +\rho_{1*}\mu_{1*}-\rho_{1*} u )]f_{U_{1*}}(u) du \\
  &=& \frac{1}{2} -\int_0^{\infty} \Phi(-\mu_{2*} +\rho_{1*}\mu_{1*}-\rho_{1*} u)f_{U_{1*}}(u)du
  \end{eqnarray*} 
  where $Z_{1*} = U_{1*}-\mu_{1*}$, $Z_{2*}=V_{1*}-\mu_{2*}$, $Z_{3*}=W_1 - \mu_{2*}+\rho_{1*}\mu_{1*}$, i.e., $Z_{1*}\sim N(0, 1)$, $Z_{2*}\sim N(0,1)$, $Z_{3*}\sim N(0, 1-\rho_{1*}^2)$ and ${\rm Cov}(Z_{1*}, Z_{3*})=0$.  
  \begin{eqnarray*}
  & &P(U_1<0, V_1<0) = P(U_1<0, W_1+\rho U_1 <0) = P(U_1<0, W_1<-\rho v) \\
  & &\;\;\;\;= \int_{-\infty}^{0} \phi(-\rho v)f_V(v) dv
  \end{eqnarray*}  
   \begin{eqnarray*}
  & &P(U_1>0, V_1<0) = P(U_1>0, W_1+\rho U_1 <0) = P(U_1>0, W_1<-\rho v) \\
  & &\;\;\;\;= \int_{0}^{\infty} \phi(-\rho v)f_V(v) dv
  \end{eqnarray*} 
   \begin{eqnarray*}
  & &P(U_1<0, V_1>0) = P(U_1<0, W_1+\rho U_1 >0) = P(U_1<0, W_1>-\rho v) \\
  & &\;\;\;\;= \int_{-\infty}^{0} [1-\phi(-\rho u)]f_U(u) du = \frac{1}{2} - \int_{-\infty}^{0} \phi(-\rho u)f_V(u)du
  \end{eqnarray*} 
  Analogously, we would have the same computations for comparisons of individuals in group 2 ($U_2$ and $V_2$) and from different groups ($U_{12}$ and $V_{12}$). 
  Therefore, 
  \begin{eqnarray*}
  & &\theta_{gg} = P({\rm Discordant}) - P({\rm Concordant}) \\
  & & \;\;\;\; = 2 \left \{ \int_0^{\infty} \Phi(-\rho v)f_V(v)dv - \int_{-\infty}^0 \phi(-\rho v)f_V(v)dv \right \} \\
   & & \;\;\;\; = 2 \int_0^{\infty} [\Phi (-\rho v) - \Phi (\rho v)] f_U(u)dv <0, \mbox{ if $\rho>0$,} \\
   & & \;\;\;\; \mbox{ since $\Phi(-\rho v)<1/2$ and  $ \Phi (\rho u)>1/2$, for $g=1, \ldots , G$}
   \end{eqnarray*}
   Within group $g$, we would have only $\rho_g$, but between groups $g$ and $g'$, we would have $\frac{\rho_1+\rho_2}{2} <\rho_2$, since $\rho_1<\rho_2$. 
   
   Note that, if $\rho_1>0$, $\rho_2>0$ and $\rho_1<\rho_2$, $2\rho_1 <\rho_1 +\rho_2 <2\rho_2$.
  \begin{eqnarray*}
  & & \theta_{11} = 2 \int_0^{\infty} [\Phi (-2\rho_1 u) - \Phi (2\rho_1 v)] f_U(u)du <0,   \\
 & & \theta_{22} = 2 \int_0^{\infty} [\Phi (-2\rho_2 u) - \Phi (2\rho_2 v)] f_U(u)du <0,   \\
  & & \theta_{12} = 2 \int_0^{\infty} [\Phi (-(\rho_1+\rho_2) u) - \Phi ((\rho_1+\rho_2) u)] f_U(u)du <0. 
   \end{eqnarray*}
   \begin{eqnarray*}
   & &  \mid \Phi (-2\rho_2 u) - \Phi (2\rho_2 u) \mid > \\
    & &\;\;\;\;\; > \mid \Phi (-(\rho_1+\rho_2) u) - \Phi ((\rho_1+\rho_2) u) \mid  > \\
    & &\;\;\;\; \; > 
   \mid \Phi (-2\rho_1 v) - \Phi (2\rho_1 v) \mid  .
   \end{eqnarray*} 
   
   Then,  $ \theta_{11} > \theta_{12} >\theta_{22}  \Rightarrow 2\theta_{12} > (\theta_{11}+\theta_{22})$.

\bibliography{sample}



\end{document}